\documentclass[11pt,a4paper]{article}
\pdfoutput=1
\usepackage{jcappub} 
\usepackage{bm}
\newcommand{\mpl}{M_{\rm P}}

\renewcommand{\d}{{\rm d}}
\newcommand{\ep}{\epsilon}

\newcommand{\fnl}{f_{\mathrm{NL}}}

\newcommand{\gnl}{g_{\mathrm{NL}}}
\newcommand{\be}{\begin{equation}}
\newcommand{\ee}{\end{equation}}
\newcommand{\beq}{\begin{equation}}
\newcommand{\eeq}{\end{equation}}
\newcommand{\bea}{\begin{eqnarray}}
\newcommand{\eea}{\end{eqnarray}}
\newcommand{\baq}{\begin{eqnarray}}
\newcommand{\eaq}{\end{eqnarray}}

\newcommand{\di}{ {\rm d}}
\graphicspath{{./}}
\notoc
 
\begin{document}

\title{Resolving primordial physics through correlated signatures}
\author{Kari Enqvist$^1$,}
\author{David J. Mulryne$^{2}$,}
\author{Sami Nurmi$^1$ }

\affiliation{\vspace{2mm}
$^1$ University of Helsinki and Helsinki Institute of Physics, P.O. Box 64, 
FI-00014, Helsinki, Finland}
\affiliation{\vspace{2mm}
$^2$ School of Physics and Astronomy, Queen Mary
University of London, Mile End Road, London, E1 4NS, UK}

\emailAdd{kari.enqvist@helsinki.fi, d.mulryne@qmul.ac.uk, sami.nurmi@helsinki.fi}

\abstract{ We discuss correlations among spectral observables as a
new tool for differentiating between models for the primordial
perturbation. We show that if generated in the isocurvature sector,
a running of the scalar spectral index is correlated with the
statistical properties of non-Gaussianities.  In particular, we find
a large running will inevitably be accompanied by a large running of
$\fnl$ and enhanced $\gnl$, with $\gnl\gg\fnl^2$. If the tensor to
scalar ratio is large, a large negative running must turn positive
on smaller scales. Interestingly, the characteristic scale of the
transition could potentially distinguish between the inflaton and
isocurvature fields.

\vspace{80mm}
}

\maketitle

\section{Introduction}

Recently an unprecedented amount of data has become available to
constrain the properties of primordial fluctuations which source
structure in our universe. This includes the results from the Planck
satellite \cite{Ade:2013zuv,Ade:2013uln,Ade:2013ydc}, the BICEP2
collaboration \cite{Ade:2014xna}, and the South Pole Telescope
(SPT)\cite{Hou:2012xq}, as well as many others. Single field slow
roll inflation, based on a scalar field, the inflaton, appears
compatible with all the data. However, in addition to the inflaton,
there might be other scalars present during inflation. Indeed,
unless the Standard Model couplings are drastically modified at
inflationary energies, we know that there is at least one light
isocurvature scalar, the Higgs, which was present during inflation
and acted as a spectator field \cite{
Enqvist:2013kaa,Espinosa:2007qp,Herranen:2014cua,Enqvist:2014bua,Fairbairn:2014zia,Kobakhidze:2013tn}.

While the SM Higgs appears to have little impact on primordial
perturbations, see however \cite{DeSimone:2012qr,DeSimone:2012gq},
isocurvature fields in general could play a significant role if an
efficient curvaton-type conversion into adiabatic perturbations
takes place. Here we take curvaton-type to cover in addition to the
curvaton scenario
\cite{Enqvist:2001zp,Lyth:2001nq,Moroi:2001ct,Linde:1996gt,
Mollerach:1989hu} also all other models where isocurvature fields
can be converted efficiently into adiabatic perturbations, e.g.
modulated reheating \cite{Dvali:2003em,Kofman:2003nx} and modulated
end of inflation \cite{Lyth:2005qk}. The Planck bound on
non-Gaussianity $f_{\rm NL}^{\rm local}=2.7 \pm 5.8$ (68\% CL) alone
places stringent constraints on curvaton-type fields as the dominant
source of primordial perturbations, see e.g.
\cite{Lyth:2002my,Sasaki:2006kq,Enqvist:2005pg,Enqvist:2009ww,Kobayashi:2013bna}.
Moreover, as is well known, this would be ruled out entirely by an
eventual detection of large amplitude primordial gravitational
waves. Constraints, however, are significantly weaker for an
admixture of curvaton and inflaton sourced perturbations
\cite{Enqvist:2013paa, Ellis:2013iea, Meyers:2013gua,
Byrnes:2014xua}. The question then arises: given the data, how one
could most efficiently constrain, or possibly detect, additional
scalars such as the curvaton?

Here we pursue one possible avenue and point out that correlations
between the observed signatures of the primordial perturbation
could provide a new handle on the dynamics during inflation.
Given that leading order effects in slow-roll are well understood,
our focus is what new can be learnt from effects at second order,
and what a detectable running of the scalar spectral index, denoted by
$\alpha$, could
tell us.
In particular, we will investigate the information encoded in the
mutual dependence of non-Gaussian statistics and the lowest order running
of the scalar spectral index.
As we will demonstrate, a large $\alpha$
would signal the presence of features in the potential, which in turn would give rise to a different
structure of non-Gaussianities for the pure inflaton and the mixed
inflaton-curvaton models, thus making it possible to distinguish between the two.

In general, there are two options for the origin of large running,
$|\alpha|={\cal O}(n_s -1)$, of the spectral index. The first lays
the responsibility with the inflaton potential. All that is required
in this case is that the inflaton's third slow-roll parameter,
$\xi$, becomes of ${\cal O} (n_s -1)$. The conventional assumption
is that $|\xi|={\cal O}( (n_s -1)^2)$. As emphasised by Stewart
\cite{Stewart:2001cd}, however, this is a superfluous assumption
often made in slow-roll inflation for reasons of convenience only.
It greatly simplifies calculations, and is satisfied by the simplest
inflationary potentials, but it is not required by general
inflationary dynamics.

The second possibility for generating a large running is that the potential
of a curvaton type isocurvature field gives rise to the running.
This option has recently been studied by Takahashi
\cite{Takahashi:2013tj} using a curvaton scenario as a concrete
example. The idea was then revisited in light of the BICEP2 data by
Sloth \cite{Sloth:2014sga}.

As we will see, the latter case leads to distinct signatures in the
non-Gaussian statistics, while in the former non-Gaussianity remains
negligible. Moreover, we find that there is a further difference
between the two cases for examples in which the running is
accompanied by a significant gravitational wave signal.
If the large running is accompanied by a significant tensor to
scalar ratio $r$, this \emph{implies that the potential contains a
feature}. This feature is not required to violate slow-roll, but
the condition required on the third slow-roll parameter cannot be
maintained for many e-folds, implying that a negative running on the pivot scale
has to be accompanied by a positive running on
shorter scales. This would correspond to a feature of a specific type
in the potential, but with differing constraints on the feature in the
inflaton and isocurvature cases.

Observational bounds on the running parameter $\alpha$ are
relatively weak. Interestingly, however, there have been several
reports of hints for a sizeable running. Analysis of the SPT data
implies a preference for $\alpha =-{\cal O} (n_s-1)$ at the
$2\sigma$ level \cite{Hou:2012xq}. More recently the BICEP2
collaboration \cite{Ade:2014xna} detected a significant B-mode
signal in CMB data, which however now looks to be mostly or even
entirely due to galactic dust \cite{Adam:2014bub}. However, if the
joint analysis of Planck and BICEP2 data currently in progress (or
any future instrument such as BICEP3, Keck, Spider 
or PIPER \cite{Ahmed:2014ixy,Sheehy:2011yf,Lazear:2014bga,Fraisse:2011xz}), 
did reveal a primordial component in the signal, the
discovery would likely favour a sizeable running of the spectral
index, see for example Refs.~\cite{Ade:2014xna, Abazajian:2014tqa}.
 In the more distant future, surveys of spectral distortions could 
significantly extend the currently accessible window of $\Delta N\sim 7$ 
inflationary e-folds. Progress
of 21-cm cosmology could extend the window even further. This would
open up entirely new possibilities to efficiently probe the scale
dependence of the spectrum.

It is therefore of great interest to carefully address
second order effects such as the running of spectral indices, and
to address the question:
\emph{ if a scenario generates a significant running, are
there further observational consequences that would result?} This is the primary
aim of the present study.

The structure of the paper is as follows. In \S~\ref{infRun} we
discuss large running generated by a feature in the inflaton
potential. This is contrasted to the isocurvature case in
\S~\ref{curvRun}, where we analyse large running from an
isocurvature field and its correlation with non-Gaussianities. In
\S~\ref{concreteIso} we present and analyse numerically two
illustrative examples of curvaton models where features in potential
generate a large running. We conclude in \S~\ref{conclusions}.

\section{Running from the inflaton field}
\label{infRun}
Let us consider first the running generated by single field inflation. In this case, for a given inflaton potential, $V(\phi)$, the tensor to scalar ratio, $r$, the spectral index, $n_s$, and
the running, $\alpha$, are given by the expressions
\baq
r &=& = 16 \epsilon \,, \nonumber \\
n_s -1 &=&-6 \epsilon+  2\eta + 1.062 \xi + \dots \,,\nonumber\\
\alpha &=&16 \epsilon \eta - 24\epsilon^2 -2 \xi  + \dots \,,
\label{srObs}
\eaq 
where the slow-roll parameters are
\baq
\epsilon &=& \mpl^2 \frac{V'^2}{2 V^2}\,, \nonumber \\
\eta &=& \mpl^2 \frac{V''}{V}\,, \nonumber \\
\xi &=& \mpl^4 \frac{V'V'''}{V^2}\,.
\eaq
The additional terms in Eq.~(\ref{srObs}) are here taken to be negligible on the
assumption that
slow-roll parameters beyond $\xi$ are subdominant compared to the first three.
In models in which $\xi$ is
of order $\epsilon$ or $\eta$, however, this is not the case in general.
A quantitative analysis then requires
a generalised slow-roll expansion
\cite{Dodelson:2001sh,Stewart:2001cd}, or a numerical analysis. The above
expressions are however sufficient for a qualitative discussion.
Considering Eqs.~(\ref{srObs}), we find that in
the absence of a fine tuned cancelation, $\epsilon$,
and $\eta$ need to be ${\cal} (n_s-1) \sim 0.01$ or much smaller, in order that
the spectral index can take a value in the observationally
preferred range $n_s=0.96 \pm0.007$ \cite{Ade:2013ydc}. It is necessary, therefore,
that $\xi$ be of a similar magnitude, if the running is also to be $ {\cal O} (n_s-1)$.

Assuming the potential supports a positive $\xi$ of this
amplitude for at least a few e-folds, the fact that $\di \eta / \di
N \sim -\xi$ implies that $\eta$ will change by at least at few times
${\cal O}(\xi)$. After the phase of evolution which generates such a
large negative running, therefore, barring fine tuned cancelations,
$\eta$ will be negative even if it was positive at the start of this
phase. Such a negative $\eta$ also feeds an increase in the rate at
which $\epsilon$ increases \emph{if $\epsilon$ is large enough for
$r$ to be significant}, through the equation $\di \epsilon/\di N = -2
\epsilon (\eta -2 \epsilon)$. The implication then is that the
inflation would not last sufficiently long if such a behaviour
continues. This will be made more transparent below. After a few
e-folds of this kind of a  behaviour, a decrease in $\epsilon$ is
needed for the potential to be able to generate a sufficient number
of e-folds of inflation.

Such a decrease requires a positive $\eta$ with $\eta > 2
\epsilon$, so the mass of the potential must switch back to
positive. This in turn requires a large negative $\xi$ and a hence a
large positive running. Again barring fine tuning, therefore,
potentials which realise a large negative running on the pivot scale
must have a large positive running on shorter scales, and must
possess an inflection point. For example, the potentials considered
by Takahashi and collaborators
\cite{Kobayashi:2010pz,Czerny:2014wua} consisted of a series of
inflection points. This behaviour implies that on shorter scales the
spectral index must evolve towards bluer values. Such behaviour is
potentially observable though joint analysis of CMB and LSS data, or
possibly on even shorter scales  by virtue of CMB spectral
distortions \cite{Clesse:2014pna} if the spectral index actually
becomes blue. The condition required for this to occur is mildly
stronger than the condition for the growth of $\epsilon$, since it
requires $\eta > 3 \epsilon$. However,  in concrete models this
condition may well be realised, potentially leading to a correlated
signature to a negative running on the pivot scale. A spectrum which
is blue tilted for an range of scales shorter is indeed what occurs
in the model of Ref.~\cite{Czerny:2014wua}.

\subsection{Explicit parametrisation of the inflaton feature}

With the preceding discussion in mind, let us therefore consider the running generated
by single field inflation
with an inflection point in the inflaton's potential.
For convenience we parameterise the potential around an
inflection point as
  \beq
  \label{Vsingle}
  V(\phi) = V_0\left(1+b \left(\frac{\phi-\phi_0}{\phi_0}\right)+c
  \left(\frac{\phi-\phi_0}{\phi_0}\right)^3+\ldots\right)\ ,
  \eeq
and we assume $b>0$ and $c>0$. At the inflection point the slow roll
parameters are given by
  \baq
  \label{e0}
  \epsilon_0&=&\frac{b^2}{2}\left(\frac{\phi_0}{\mpl}\right)^{-2}\ ,\\
  \label{eta0}
  \eta_0&=&0\ ,\\
  \label{xi0}
  \xi_0&=&6cb\left(\frac{\phi_0}{\mpl}\right)^{-4}\, .
  \eaq

Here we concentrate on the case $\xi_0\gg \epsilon_0^2$ where a
large negative running $\alpha_0 \simeq -2\xi$ is generated at the
inflection point. Assuming the higher order terms in (\ref{Vsingle})
are negligible, the slow roll inflation can be sustained only for a
limited range of e-folds
  \beq
  N_{\rm max}=\frac{1}{\sqrt{|\alpha_0|}}\left(2 {\rm arctan}(x_{\rm e}) -\frac{r_0}{12\sqrt{|\alpha_0}|}
  \left(x_{\rm e}^2+2{\rm ln}(1+x_{\rm e}^2)\right)\right)\ , \qquad x_{\rm e} =\sqrt{\frac{3c}{b}}\,\frac{\phi_0-\phi_{\rm e}}{\phi_0}
  \eeq
In the limit of large gravitational waves $r_0 > 4|\alpha_0|$ the
slow roll ends as $\epsilon(\phi_e)=1$ and the corresponding field
value is determined by the condition
  \beq
  x_{\rm e}^2+\sqrt{\frac{r_0}{|\alpha_0|}}\left(x_{\rm e}+\frac{1}{3}x_{\rm e}^3\right)
  = \frac{4}{\sqrt{r_0}}-1\ .
  \eeq
For $r_0=0.2$ and $\alpha_{0}=0.02$ this yields $N_{\rm max}\lesssim
10$. Decreasing the running and the tensor-to-scalar ratio to
$r_0=0.1$ and $\alpha_{0}=0.01$ increases the number of e-folds up
to $N_{\rm max}\lesssim 15$.
 
Therefore, if $r={\cal O}(0.1)$ and $\alpha = {\cal O}(n_s-1)$ at
the horizon crossing of observable modes the potential
(\ref{Vsingle}) necessarily needs to be flattened out by higher
order terms within $N = {\cal O}(10)$ e-folds. The flattening
essentially requires a second inflection point and a subsequent
evolution of $\eta$ to positive values. This will push the spectral
index $n_{s}$ towards, and possibly over, to blue tilted values, as
discussed above.

It would be interesting to investigate how generically blue values
will be reached, and if this effect, which necessarily follows from
the large running and large tensor-to-scalar ratio in single field
models of the form (\ref{Vsingle}), could be detectable through
future observations of LSS, CMB spectral distortions or 21-cm
cosmology. However, we defer a detailed study to future work.
Examples of this behaviour in models with a large running can be
found in for example Refs.~\cite{ Dodelson:2001sh,Czerny:2014wua}


\section{Running from an isocurvature field}
\label{curvRun}

Let us now consider the case where more than one light field is
present during inflation. In this case the running can be generated
by isocurvature fields which after the end of inflation convert
their fluctuations into adiabatic perturbations.

Using the $\delta N$ formalism \cite{Lyth:1984gv,
Starobinsky:1986fxa, Wands:2000dp, Sasaki:1995aw, Lyth:2005fi} the
spectral index to lowest order in slow roll is given by
  \be n_s-1 = -2
\epsilon^* - 2\frac{1-\eta_{ab}N_a N_b}{N_c N_c}+\dots\ .
\label{nsIso}
  \ee
Here $N_a$ are the derivatives of $N$ defined in the usual way
\cite{Lyth:2005fi}, and the subscript Roman letters run over all
light fields present. In a similar manner to the single field case,
however, additional terms in Eq.~(\ref{nsIso}) cannot in general be
neglected if the running is significant. Here we will use
expressions such Eq.~(\ref{nsIso}) only to guide us in analytical
estimates, but where necessary use numerical analysis which does not
rely on them.

Differentiating this expression with respect to $\ln k = \ln(aH)$
provides an expression for the running, and one finds \be \frac{\d
n_s}{\d \ln k} \supseteq -\frac{2}{N_f N_f} \frac{ \xi^*_{ac} N_a
N_c}{N_f N_f} + \dots \ee where $\xi_{ab} = V_{abc} V_c/V^2$. As in
the single field case, to enhance running while adhering to
slow-roll typically requires $\xi = {\cal O}(\epsilon)$, while in
the simplest models $\xi = {\cal O}(\epsilon^2)$.
 In this section we consider two field models, and the
consequences of a large $\xi_{\chi \chi}$, where $\chi$ is considered
to be a field which is a spectator at horizon crossing. This will lead to a large running if the perturbation in this field is converted into
a curvature perturbation. We now discuss the conditions needed for the
enhancement of the running, as well as the number
of e-folds over which such an enhancement can be maintained.

\subsection{Explicit parametrisation of the isocurvature feature}

In a manner similar to the inflaton case above, we consider a
potential for the isocurvaure field $\chi$ expanded about an
inflection point, given by
\beq
\label{Vchi}
U(\chi) = U_0\left(1+ \beta \left(\frac{\chi-\chi_0}{\chi_0}\right)
+ \gamma \left(\frac{\chi-\chi_0}{\chi_0}\right)^3 + \ldots\right)\
.
\eeq
Here $U_0,\beta$ and $\gamma$ are constants and we take the full
scalar potential to be of the form $V=V(\phi)+U(\chi)$.

In order to be an isocurvature field, $\chi$ should be energetically
subdominant and the gradient of the full potential in $\chi$
direction should vanish to first order in slow roll
  \beq
  \label{epsilonchiconst}
  \Omega_{\chi} = \frac{U}{3H^2M_{\rm P}^2}\ll 1\ ,\qquad \epsilon_{\chi}\lesssim \epsilon_{H}^2\ .
  \eeq

A conversion of the isocurvature fluctuations of $\chi$ into
adiabatic perturbations after inflation could generate a sizeable
running of the spectral index, as suggested by Takahashi \cite{Takahashi:2013tj} and Sloth \cite{Sloth:2014sga}. In the examples considered here, the eventual magnitude of the running is controlled
by the $c$ in the potential (\ref{Vchi}) expanded around the
inflection point. Assuming the third derivative of the inflaton
potential is negligible, the running generated through the
isocurvature field at scales exiting the horizon as $\chi=\chi_0$ is
given by
 \beq
  \alpha_0 = -2 w_{\chi} \xi_{\chi\chi} + {\cal O}(\epsilon_{H}^2)\ ,\qquad \xi_{\chi\chi}=12\, b\, c\,  \Omega_{\chi}^2
  \left(\frac{M_P}{\chi_0}\right)^4\,.
  \eeq
Here $w_{\chi} = {\cal P}_{\chi}/ {\cal P}_{\zeta}$ is the $\chi$
contribution to the total curvature perturbation at the time when
the system becomes adiabatic and $\zeta$ freezes to constant. For a
large tensor perturbation with $r\sim 0.1$, the isocurvature
contribution could not account for the total curvature perturbation
but its typical contribution is constrained by $w_{\chi}\lesssim
0.5$ \cite{Enqvist:2013paa,Byrnes:2014xua}.

The range of $\chi_0$ values for which the isocurvature conditions
(\ref{epsilonchiconst}) are satisfied and a large running
$|\alpha_0|\gtrsim \epsilon_{\rm H}$ can be obtained at least
locally is given by
  \beq
  \label{chi0bigger}
  b \left(\frac{\Omega_{\chi}}{
  10^{-7}}\right)\left(\frac{0.1}{r}\right)^{3/2}\lesssim
  \frac{\chi_0}{H}\lesssim \left(\frac{\Omega_{\chi}}{
  10^{-5}}\right)^{1/2} w_{\chi}^{1/4} b^{1/4} \left( \frac{c}{
  10^{-5}}\right)^{1/4} \left(\frac{0.1}{r}\right)^{3/4}\ .
  \eeq

Moving away from the inflection point $\epsilon_{\chi}$ will rapidly
grow due to the large running. Unless the growth is leveled out by
higher order terms in (\ref{Vchi}) the isocurvature condition
(\ref{epsilonchiconst}) eventually gets violated as
  \beq
  \epsilon_{\chi} =
  \epsilon_{\chi_0}\left(1+{\rm
  tan}^2\left(\frac{\xi_{\chi\chi}}{6} N \right)\right)^2
  \simeq  \epsilon_{H}^2\ .
  \eeq
The corresponding maximum amount of e-folds after the inflection
point under which $\chi$ remains an isocurvature field is given by
  \beq
  N_{\rm max}\simeq \sqrt{\frac{4 w_{\chi}}{|\alpha_0|}}{\rm
  arctan}\left(\sqrt{\frac{\epsilon_H^2}{\epsilon_{\chi_0}}-1}\right)
  <  \pi\sqrt{\frac{w_{\chi}}{|\alpha_0|}}
  \eeq
Taking $\alpha_0\sim 0.01$ with $w_{\chi}\lesssim 0.5 $ as a
representative example we find
  \beq
  N_{\rm max} \lesssim 30\ ,
  \eeq
no matter how small the isocurvature slope $\epsilon_{\chi_0}$ would
be at the inflection point. In order to keep $\chi$ as an
isocurvature field, its potential necessarily needs to be levelled
out by higher order terms in (\ref{Vchi}) within the $N\sim 30$
e-folds after the inflection point.

We therefore find that a single feature in the isocurvature
potential is in general not enough to generate a large running
$|\alpha|\sim 0.01$ while still keeping the isocurvature field
decoupled from the adiabatic direction over the observable $N_{\rm
CMB} \sim 60$ e-folds. It appears that a large running from an
isocurvature field can be consistently generated only if there are
multiple features in the isocurvature potential, as was the case for
the inflaton potential. In both setups a large negative running
$\alpha=-{\cal O}(n_s-1)$ needs to be followed by a transition to
positive running on smaller scales.

The characteristic scale of the transition however distinguishes the
inflaton induced running from the isocurvature case. In the inflaton
case the transition from positive to negative running occurs within
$N\sim 10$ e-folds after the inflection point whereas for the
isocurvature case the transition can take place at much smaller
scales. This difference could be observable by future measurements
of spectral distortions and 21-cm spectrum which are expected to
significantly extend the currently observable range $\Delta N \sim
7$ of CMB scales.

\subsection{Estimates of isocurvature features in $\fnl$ and $\gnl$ }

We now consider further
consequences of a large running generated in the
isocurvature sector. In particular, we will consider the
amplitude of the reduced bispectrum $\fnl$, and of
trispectrum parameter $\gnl$, given respectively  by
\be
\fnl = \frac{5}{6}\frac{N_a N_b N_{ab} }{(N_c N_c)^2}
\ee
and
\be
\gnl =  \frac{25}{54}\frac{N_a N_b N_c N_{abc}}{(N_d N_d)^3}~.
\ee

In a multi-field inflationary model, the derivatives of $N$ are not unrelated to one another. Rather one finds \cite{Sasaki:1995aw}
\be
\frac{N_a V_a}{V} = 1
\ee
and differentiating this expression
\be
\label{eq:rel2}
\frac{N_{ab}V_a}{V} = \frac{N_aV_aV_b}{V^2} - \frac{N_a V_{ab}}{V}
\ee
and
\be
\label{eq:rel3}
\frac{N_{abc} V_a}{V} + \frac{N_{ab} V_{ac}}{V} + \frac{N_{ac} V_{ac}}{V} + \frac{N_aV_{abc}}{V} = \frac{V_{bc}}{V} \,.
\ee

In a two field model in which one field is a spectator at horizon
crossing, the derivative of $N$ for the inflation field $\phi$ is
very nearly constant from horizon crossing for the entire subsequent
evolution, and is given by $|N_\phi| \approx 1/(2
\epsilon^*_{\phi})^{1/2}$ (with $M_{\rm P}=1$). On the other hand,
$N_\chi$ will grow from a negligible initial value if isocurvature
fluctuations are converted into $\zeta$.

We are interested in the case where $N_\chi$ does indeed grow, and moreover, in order to make analytic progress we will often consider the point at which $N_\chi\approx N_\phi$. This condition is reasonable in our study to gain an estimate
of the effects of a large running, since we know the isocurvture field
must contribute significantly to $\zeta$ for the running of $\zeta$ to be significant, while
if there is a significant detection of gravitational waves,
it must not dominate it.  If $N_\chi$ came to dominate
$r$ would become negligible since
\be
r = \frac{16 \ep^*}{1+R} \,, \label{eq:r}
\ee
where $R=N_{\chi}^2/N_{\phi}^2$. Moreover, the condition $R=1$ typically
coincides approximately with the peak value of $\fnl$ generated
during the evolution of $\zeta$ in a two-field model \cite{Elliston:2011dr,Elliston:2013afa}. We see numerically that
our conclusions regarding non-Gaussianity hold more generally than for the case
$N_\chi \approx N_\phi$, and thus do not require a significant value of $r$.

Considering Eq.~(\ref{eq:rel2}), and assuming $V_{\phi \chi} = 0$, we find
\be
\label{large1}
N_{\phi \chi} V_\phi + N_{\chi \chi} V_\chi = V_\chi - N_\chi
V_{\chi \chi} \ee assuming there is no cancelation between the two
terms on the lhs, when $\fnl$ is large, its value when $N_\phi =
N_\chi$, which is close to it peak value can be estimated by \be
\frac{6}{5}\fnl \approx \frac{1}{2}{N_{\chi \chi}}\epsilon^{*}_\phi
= {\cal O} \left ( -\frac{\eta_{\chi \chi}}{4}
\left(\frac{\epsilon^{*}_\phi}{\epsilon^{*}_\chi} \right )^{1/2}
\right )\,. \ee Playing the same game with Eq.~(\ref{eq:rel3}), and
again assuming there is no cancellation one finds that when $\xi$ is
first order in slow-roll, as required to source a large running,
then the leading term gives \be \frac{54}{25}\gnl \approx
\frac{1}{2^{3/2}} N_{\chi \chi \chi}{\epsilon^{*}}^{3/2} = {\cal O}
\left (
\frac{-\xi^*_{\chi\chi}}{8}\frac{\epsilon^{*}_\phi}{\epsilon^{*}_\chi}
\right) \ee and hence \be \gnl = {\cal O}\left ( -\fnl^2
\frac{\xi_{\chi \chi}}{\eta_{\chi \chi}^2}\right)\,. \ee The last
expression is clearly only valid if $\eta_{\chi \chi}$ is non-zero.
Consequently we see that when a large running is present we expect
an enhanced value of $\gnl$. A further consequence is a large
running
\cite{Sefusatti:2009xu,Byrnes:2010ft,Byrnes:2009pe,Byrnes:2009qy,Shandera:2010ei}
of $\fnl$. This is because in the same approximation \be \frac{\di
\fnl}{\di \ln k} \supseteq -\frac{5}{6} \gnl
\frac{\sqrt{\epsilon_\chi}}{\sqrt{\epsilon_\phi}}\,. \ee As we will
see this enhanced running will prove to be a constraint on otherwise
viable models.

A word of caution about these expressions is in order. There are two terms on the LHS of the expressions above we use to generate our estimates for $\fnl$ and $\gnl$, and either or both of these can contribute towards any large value
on the RHS. This means that the expressions we
write down are only
order of magnitude considerations, and
cancellations between terms can render them inaccurate, even altering the sign of $\fnl$ and $\gnl$. Applying the
$\fnl$ expression above to the simplest curvaton model with a quadratic curvaton potential, for example, gives an estimate for $\fnl$ of the correct magnitude, but gives the incorrect sign.

Below we will see that these estimates work very well for two separate models which generate a large running (for both sign and magnitude). Never-the-less they should be regarded as useful guides rather than concrete expressions, and we will use numerical $\delta N$ simulations to confirm our expectations in each case.

\section{Concrete isocurvature examples}
\label{concreteIso}

To provide concrete examples of models of inflation in which a large
running is generated in the isocurvature sector together with a
correlated signal in the bispectrum and trispectrum, we consider the
curvaton model
\cite{Mollerach:1989hu,Linde:1996gt,Enqvist:2001zp,Lyth:2001nq,
Moroi:2001ct,Enqvist:2005pg,Linde:2005yw,Malik:2006pm,Sasaki:2006kq,
Meyers:2013gua, Elliston:2014zea}. This is a two field model in
which one field, which is a spectator at horizon crossing decays to
radiation long after the other field which dominates the energy
density during inflation. During the oscillations of this field its
relative contribution to the energy density can gradually increase,
and so too does the contribution of its perturbations to the
curvature perturbation $\zeta$. We restrict our attention to this
model, although similar results would be expected in other two-field
models, for example those which generate a large $\gnl$ during
inflation when $\xi_{\chi \chi}$ is large \cite{Elliston:2012wm}.

\subsection{Oscillatory isocurvature potential}

\begin{figure}
\centering
{
\mbox{\includegraphics[width=0.4\textwidth]{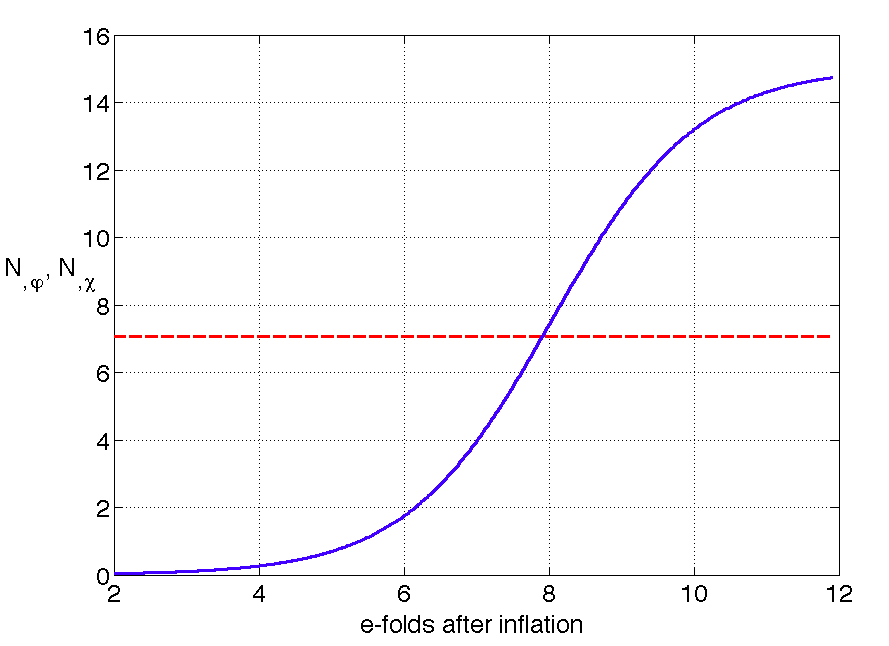}
\includegraphics[width=0.4\textwidth]{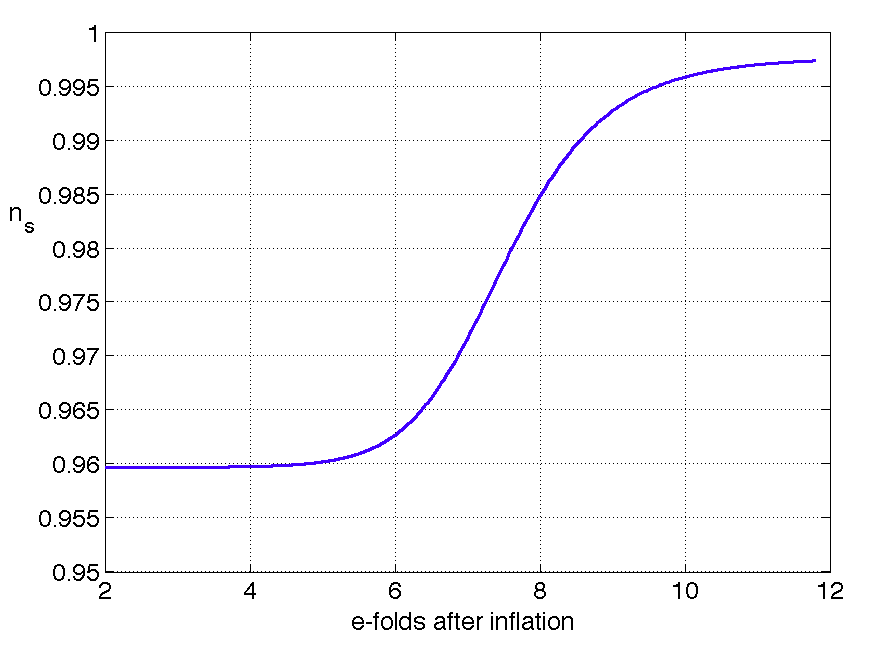} }
\mbox{\includegraphics[width=0.4\textwidth]{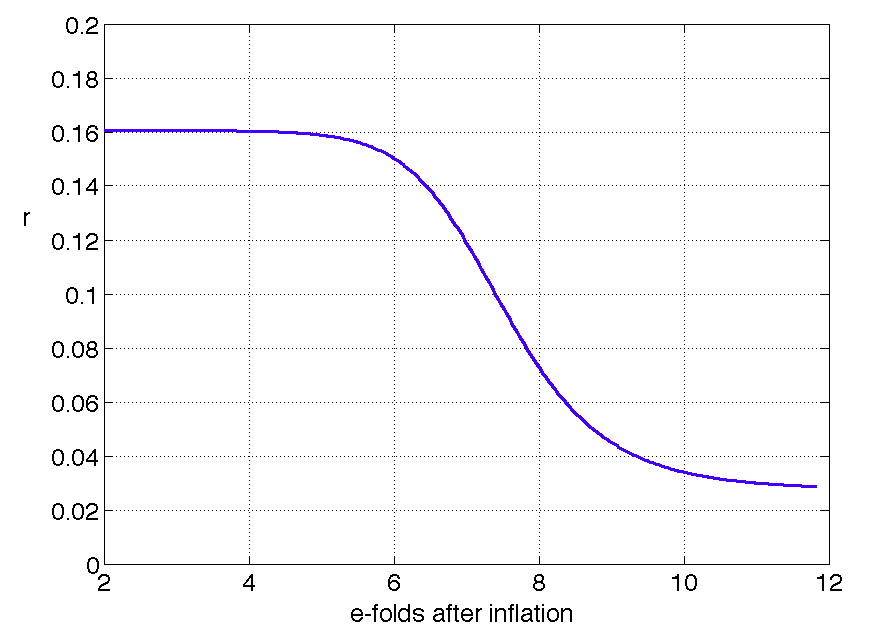}
\includegraphics[width=0.4\textwidth]{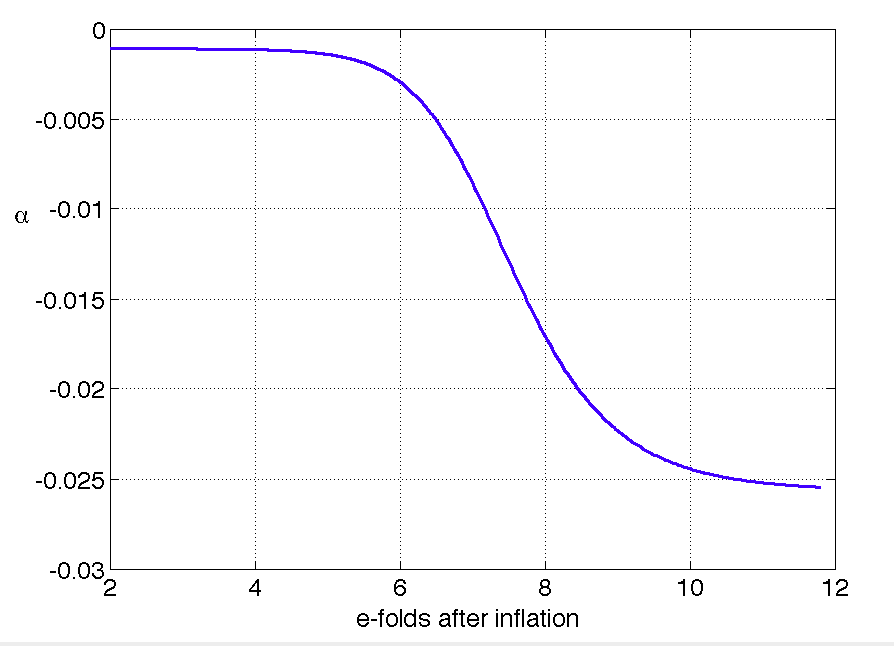} }}
\mbox{\includegraphics[width=0.4\textwidth]{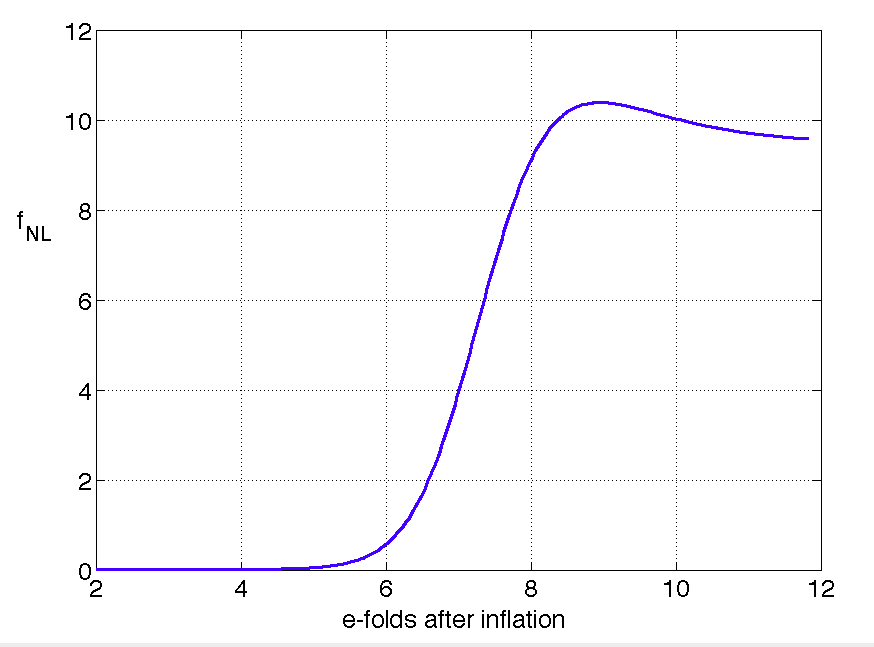}
\includegraphics[width=0.4\textwidth]{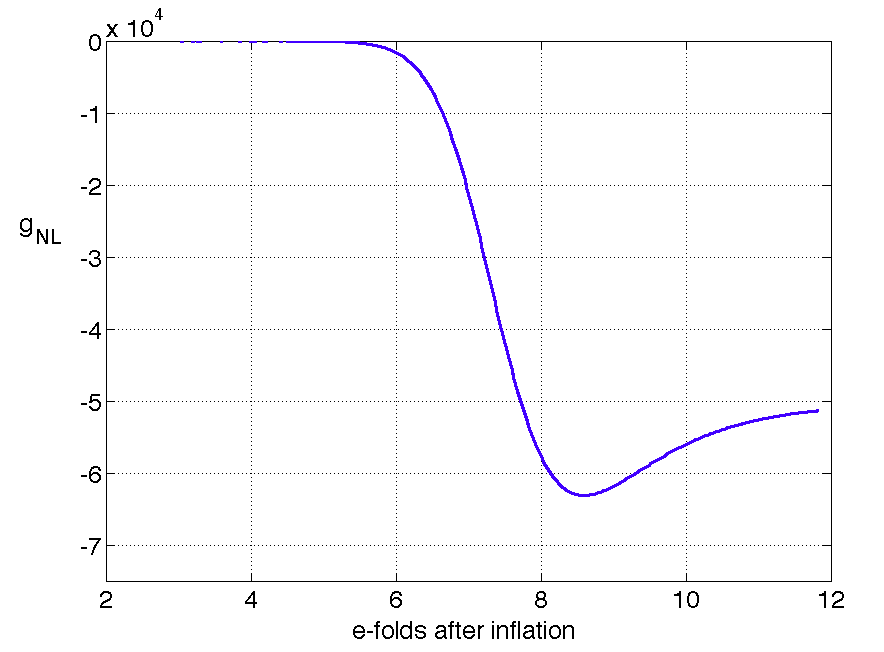} }
\caption{Results for the oscillatory isocurvature potential eq.
(\protect\ref{osc_pot}) with $m_\phi = 5 m_\chi$, $A = 0.0001$ and $
w=4\times10^{-5}$ and initial conditions $\phi = 14\mpl$ and  $\chi
= w(1/2+2*198)\pi \mpl$ set $N=50$ e-folds before the end of
inflation. The figures present the time evolution (in e-folds) from
the end of inflation of $N_\phi$ (dashed line top left), $N_\chi$
(solid line top left), $n_s$ (top right), $r$ (middle left),
$\alpha$ (middle right), $\fnl$ (bottom left) and $\gnl$ (bottom
right).
\label{fig1}}
\end{figure}

\begin{figure}
\centering { \mbox{\includegraphics[width=0.4\textwidth]{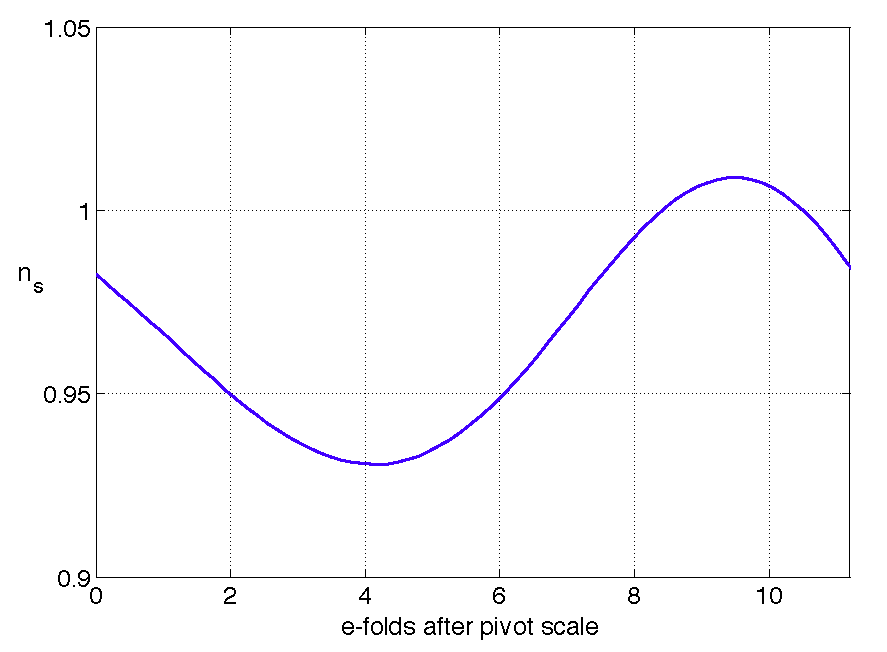}
\includegraphics[width=0.4\textwidth]{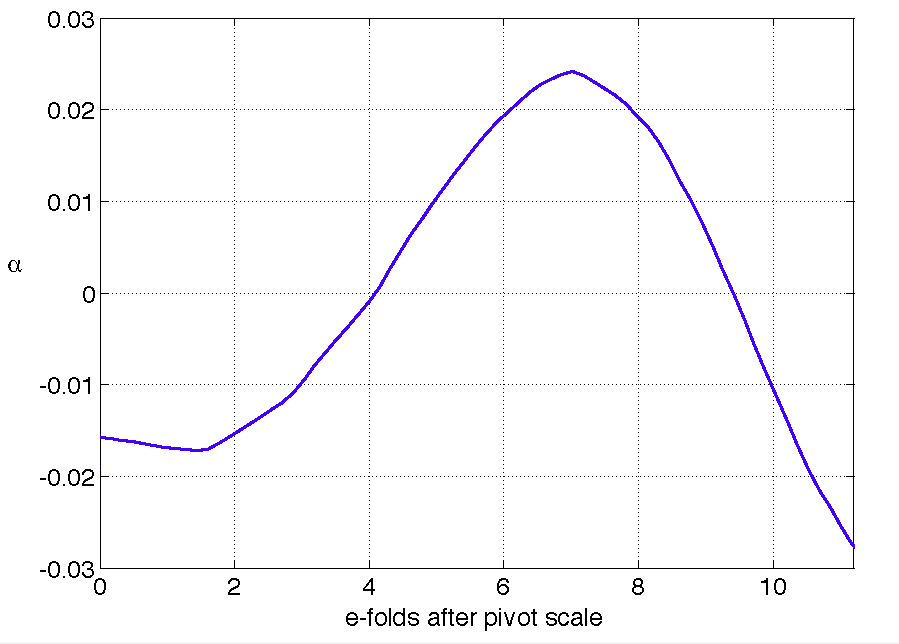} }
\mbox{\includegraphics[width=0.4\textwidth]{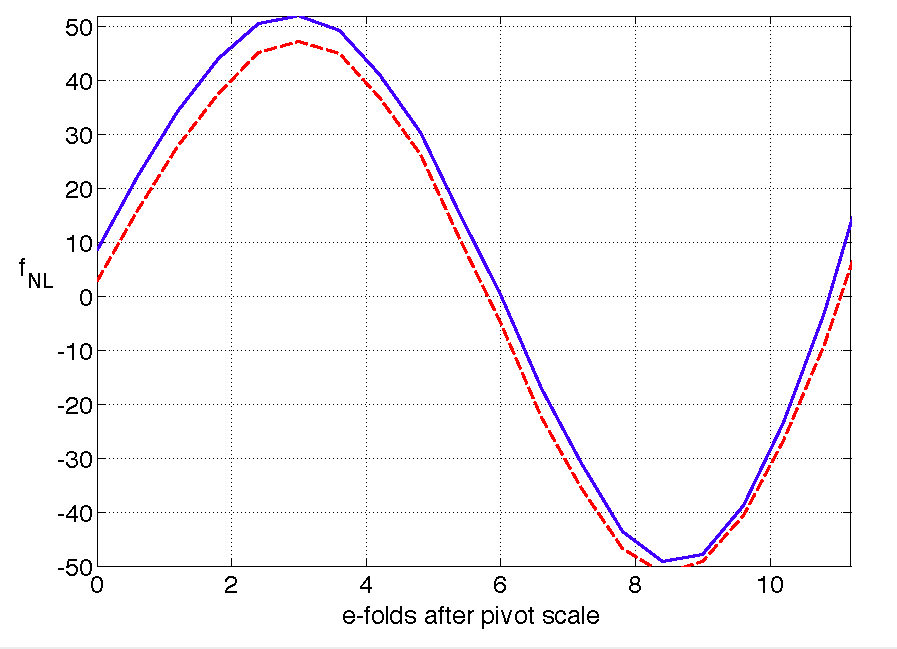}
\includegraphics[width=0.4\textwidth]{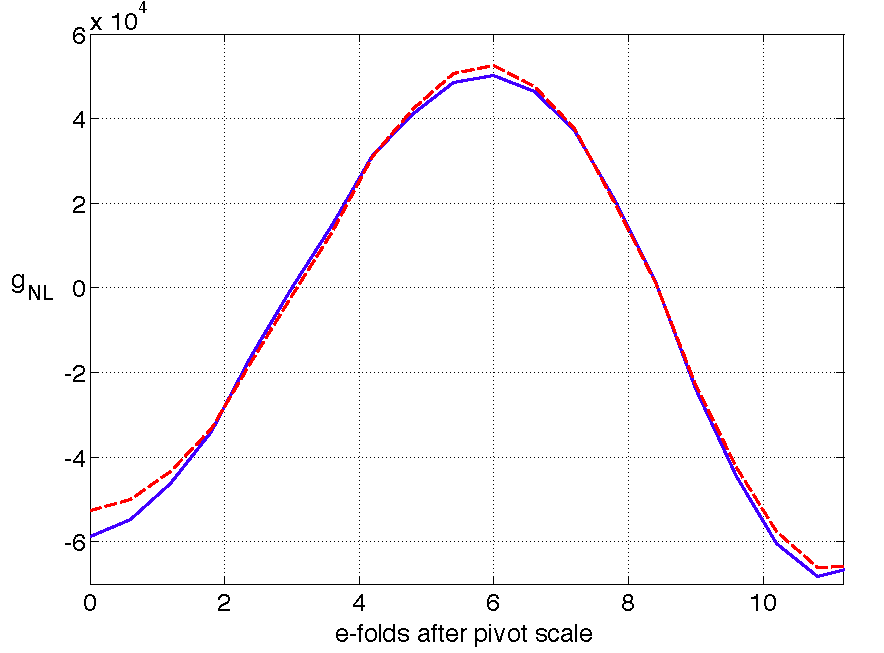} }
\caption{Results for the oscillatory isocurvature potential eq.
(\protect\ref{osc_pot}) with $m_\phi = 5 m_\chi$, $A = 0.0001$ and $
w=4\times10^{-5}$ and initial conditions $\phi = 14\mpl$ and  $\chi
= w(1/2+2*198)\pi \mpl$ set $N=50$ e-folds before the end of
inflation. The figures present the scale dependence of $n_s$ (top
left), $r$ (top right), $\fnl$ (bottom left solid), the estimate of
$\fnl$ (bottom right dashed), $\gnl$ (bottom right solid) and the
estimate of $\gnl$ (bottom right dashed). The results are shown as a
function of e-folds from the pivot scale $N=50$ and measured at the
point $N_\phi =N_\chi$ for the pivot scale.}
\label{fig2}}
\end{figure}

As a first a concrete example, therefore, we follow Takahashi and Sloth by considering a curvaton model where the curvaton potential has subdominant oscillations.
The potential we use is given by
\begin{eqnarray}
\label{osc_pot}
V(\phi)& =& \frac{1}{2} m_\phi^2 \phi^2 \,, \nonumber \\
V(\chi) &=& \frac{1}{2} m_\chi^2 \chi^2 \left ( 1 + A \cos (\chi/w) \right )\,,
\end{eqnarray}
where $\phi$ is the inflation and $\chi$ the curvaton, and we have chosen a slightly different potential to Takahashi
so as to ensure the mass at the minimum of the curvaton potential
is given by $m_\chi$.

We take $A<1$, and for $w<1$ find that potential
contains successive inflection points when  $\chi \approx w (1/2+2n) \pi$ and $\chi \approx w (3/2+2n) \pi$. The
condition that the field does not get stuck in a local maxima requires ${\chi^*}< w/(2A)$.
At the inflection points one finds
\be
\xi_{\chi \chi} \approx \left(1 \pm \frac{A \chi}{2 w} \right) \frac{A \chi^3 m^4_\chi}{2 w^3 V(\phi)^2} \,.
\ee
and that increasing $A$ or $m_\phi$,
or decreasing $\omega$ increases $\xi$.

\subsection{Numerical evolution}

As a concrete example we take the parameters values: $m_\phi = 5
m_\chi$, $A = 0.0001$ and $ w=4\times10^{-5}$. Together with the
field values when scales corresponding to the pivot scale exited the
horizon (which we take to be 50 e-folds before the end of
inflation): $\phi = 14\mpl$ and  $\chi = w(1/2+2*198)\pi \mpl$. In
this example we are not being overly careful to ensure consistency
with observation, but rather aim to illustrate the features we have
been discussing for models with a significant running and
significant production of gravitational waves. To generate our
results we numerically evolve the covariance matrix $\langle \delta
\phi_a(k) \delta \phi_b(k)\rangle$ until all scales we are interested in
have exited the horizon, the method we use is described in detail
elsewhere \cite{Ellis:2014rja}. Then we relate this to the spectrum
for curvature perturbation $\zeta$ by calculating the derivatives of
$N$ from the point we stop evolving the covariance matrix. To
calculate non-Gaussianities, we use calculate the derivatives of $N$
over a range of scales, from the pivot scale down to the final scale
we evolve the covariance matrix for. We use the numerical method
discussed in Ref.~\cite{Elliston:2011dr}, with the added assumptions
that the inflaton reheats instantaneously when it starts to
oscillate, and that once the curvaton has undergone a large number
of oscillations it may be approximated by a fluid with zero
pressure.

In Fig.~\ref{fig1}, we show how $N_\phi$, $N_\chi$ calculated on the
pivot scale, $n_s$ $r$, $\alpha$, $\fnl$, and $\gnl$ evolve during
the phase of evolution in which the curvaton is oscillating about
its minimum and its energy density red-shifting more slowly than
that of the radiation produced by the inflaton field. In these
figures different numbers of e-folds along the $x$ axis correspond
to different reheating times for the curvaton field, and hence
different energy scales for the reheating. In a real model the value
of the observational parameters which are realised is then fixed by
the curvaton's reheating scale. Plotting the results in this way
allows us to simultaneously present results for many different
reheating scales at once.

In these figures
one can see the almost
constant $N_\phi$ described above,
the growing $N_\chi$, and the generation of the large
running in the spectral index as the curvaton starts
to contaminate the spectrum.
We see in this example that
at a reheating time of about $7$ e-folds after inflation
ends $n_s$ is close to the observationally preferred value, and $r\approx 0.1$.
 As expected,
$\alpha$ is enhanced. It is clear that there is an very enhanced value of
$\gnl$ which accompanies an enhanced $\alpha$, irrespective of the value of $r$.

Next we plot how the observables behave as a function of horizon
exit time. To do this we calculate $n_s$, $\alpha$, $\fnl$ and
$\gnl$ as a function of horizon exit time over the range of e-fold
taken for the isocurvature field to evolve over on oscillation in
the potential. We evaluate the answer at the point where the
derivatives of $N$ are equal for the pivot scale. We expect the
estimates for $\fnl$ and $\gnl$ above to be accurate to at least at
the order of magnitude level at this point. We plot the results in
Fig.~\ref{fig2}. This again clearly illustrates the correlation
between $\gnl$ and the running of the spectral index for a variety
of initial conditions, in addition to the usefulness of the
estimates we have derived.

\begin{figure}[htb]
\centering
{
\mbox{\includegraphics[width=0.4\textwidth]{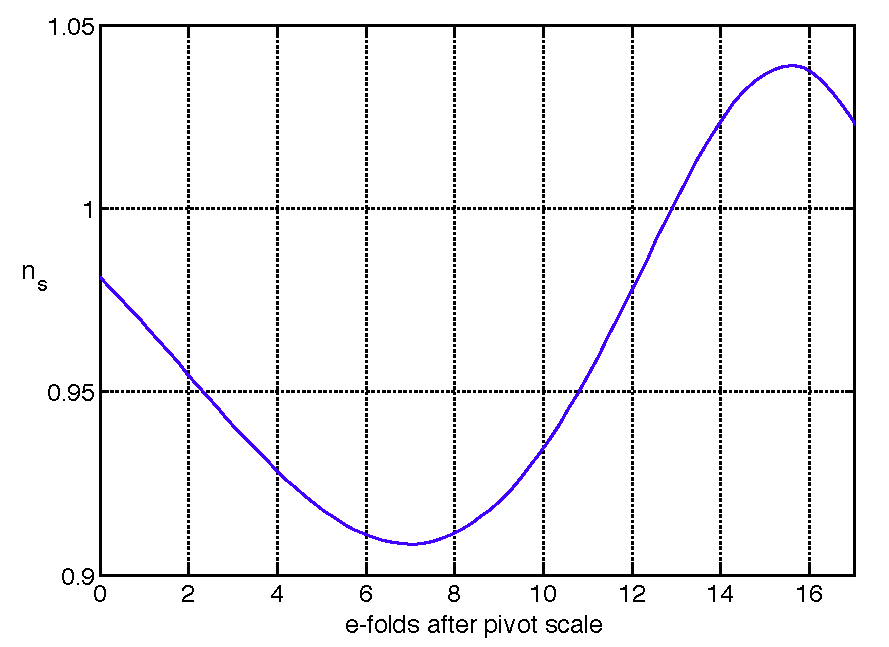}
\includegraphics[width=0.4\textwidth]{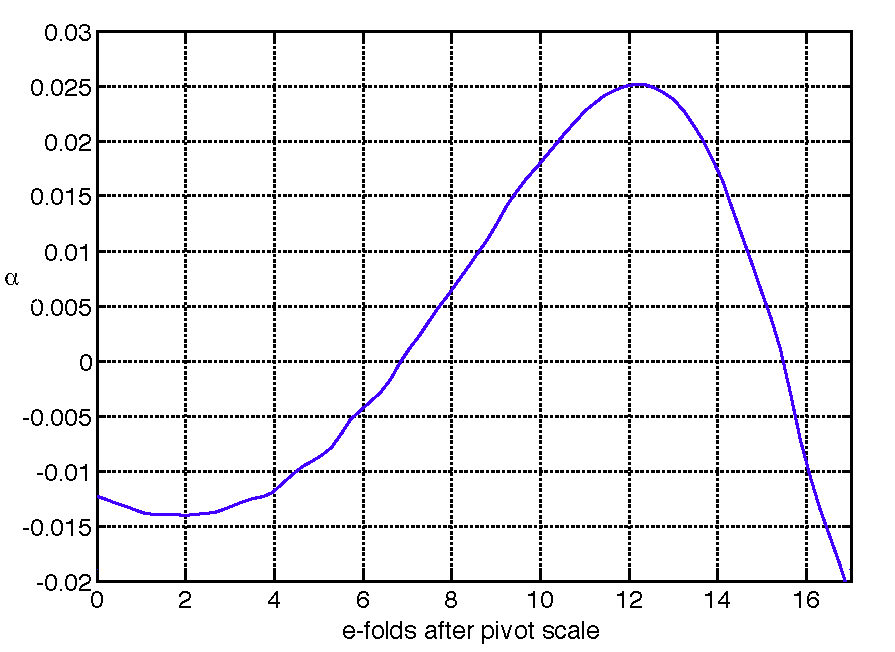} }
\mbox{\includegraphics[width=0.4\textwidth]{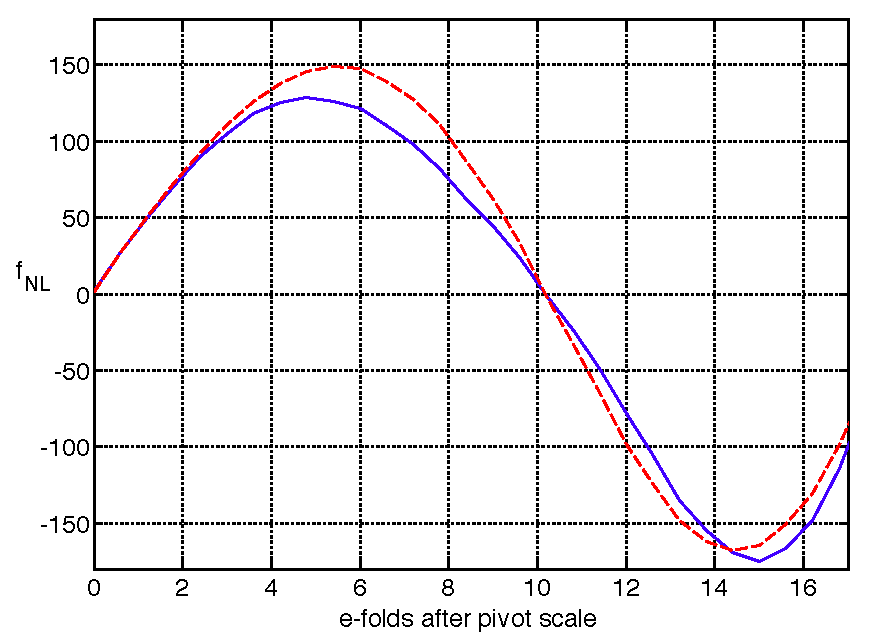}
\includegraphics[width=0.4\textwidth]{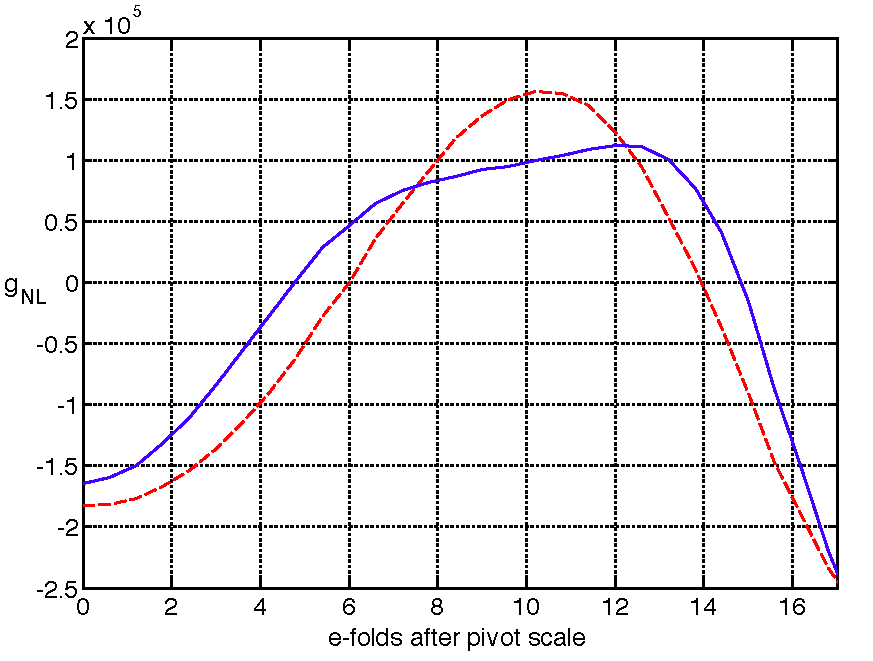} }
\caption{ Results for the oscillatory isocurvature potential eq.
(\protect\ref{osc_pot}) with $m_\phi = 20/3 m_\chi$, $A = 0.0003$
and $ w=4\times10^{-5}$ and initial conditions $\phi = 14\mpl$ and
$\chi = w(1/2+2*198)\pi \mpl$ set $N=50$ e-folds before the end of
inflation. The figures present the scale dependence of $n_s$ (top
left), $r$ (top right), $\fnl$ (bottom left solid), the estimate of
$\fnl$ (bottom right dashed), $\gnl$ (bottom right solid) and the
estimate of $\gnl$ (bottom right dashed), as a function of e-folds
from the pivot scale fixed at $N=50$ and measured at the point
$N_\phi =N_\chi$ for the pivot scale.}
\label{fig3}}
\end{figure}

In  Fig.~\ref{fig3}, we repeat this exercise but for the case
$m_\phi = 20/3 m_\chi$, $A = 0.0003$ and $ w=4\times10^{-5}$, and the same initial conditions.
In this case the values of $n_s$ and $\alpha$ are similar at horizon
crossing, but the slower rate at which the isocurvature field evolves means it takes roughly
twice as long in e-folds for on oscillation of the potential to be traversed. In this case
the large running of $\fnl$, and the extra e-folds over which it applies, means that
$\fnl$ becomes large over CMB scales, and this model is likely ruled out.

\begin{figure}[htb]
\centering {
\mbox{\includegraphics[width=0.6\textwidth]{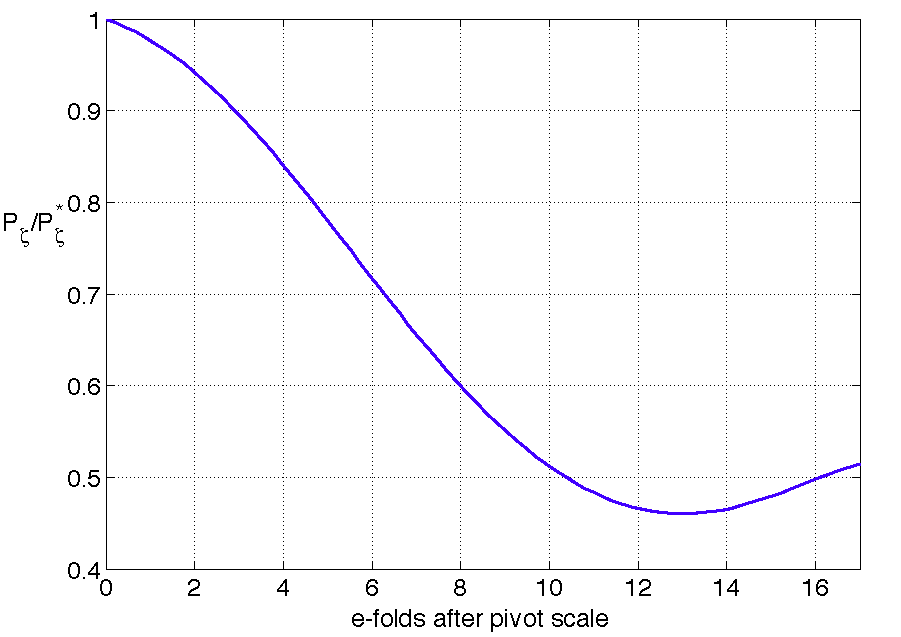} }
\caption{Results for the oscillatory isocurvature potential eq.
(\protect\ref{osc_pot}) with $m_\phi = 20/3 m_\chi$, $A = 0.0003$
and $ w=4\times10^{-5}$ and initial conditions $\phi = 14\mpl$ and
$\chi = w(1/2+2*198)\pi \mpl$ set $N=50$ e-folds before the end of
inflation. A plot of the spectrum over one oscillation as a function
of e-folds from pivot scale $N=50$.}
\label{fig4}}
\end{figure}

Finally for this case we plot the spectra of $\zeta$ directly as a function of e-folds
after horizon crossing in Fig.\ref{fig4}.

\subsection{Piecewise isocurvature potential}
As a second example, in order to have more
control over the shape of the potential,
we study an isolated feature
rather the a continuous series as in the previous example.
We therefore consider a potential of the form (\ref{Vchi}),
including also a quartic piece in the potential, and
match it to a potential of purely quadratic form about the
minimum, i.e. $V=1/2 m_\chi^2\chi^2$. We do so by requiring that the first and second derivatives of the potential agree
at some matching position $\chi_m$. This fixes the coefficient of the
quartic term, and $V_0$, leaving us free to adjust
$b$ and $c$ as we wish. In particular, this allows us to
arbitrary fix $\xi = \mpl^4 b c/ \chi_0^4$.
Explicitly the potential is then of the from
\begin{eqnarray}
\label{piecewise}
V(\phi) &=& \frac{1}{2} m_\phi^2 \phi^2 \,, \nonumber \\
V(\chi) &=& V_0 \left( 1 + b \left( \frac{\chi-\chi_0}{\chi_0} \right)  +
c \left ( \frac{\chi-\chi_0}{\chi_0} \right)^3 +
d \left ( \frac{\chi-\chi_0}{\chi_0} \right )^4  \right )\,, ~~~~\chi>\chi_m \nonumber \\
  V(\chi) &=& \frac{1}{2} m_\chi^2 \chi^2\,, ~~~~ \chi <\chi_m \,.
\end{eqnarray}

In this case as a concrete example we
generate observable results evaluated on the pivot scale with initial
conditions
$\phi^* = 14 \mpl$, $\chi_* = \chi_0 = 0.05 \mpl$,
and parameter values $m_\phi = 5 m_\chi$,
$V_0 b/\chi_0=m_\phi^2 \chi_0$  and $\chi_m = 0.049\mpl$.
The initial value of $\chi$ is therefore
at the inflection point.
We do so for a series of values of $V_0 c/\chi_0^3$
in the range $\{0, 0.01 \}$. In contrast with the example
above, therefore, we enforce $50$ e-folds for a series
of different model parameters and plot how the observables
evolve with the parameters, rather than as a function of scale.

\begin{figure}[htb]
\centering
{
\mbox{\includegraphics[width=0.44\textwidth]{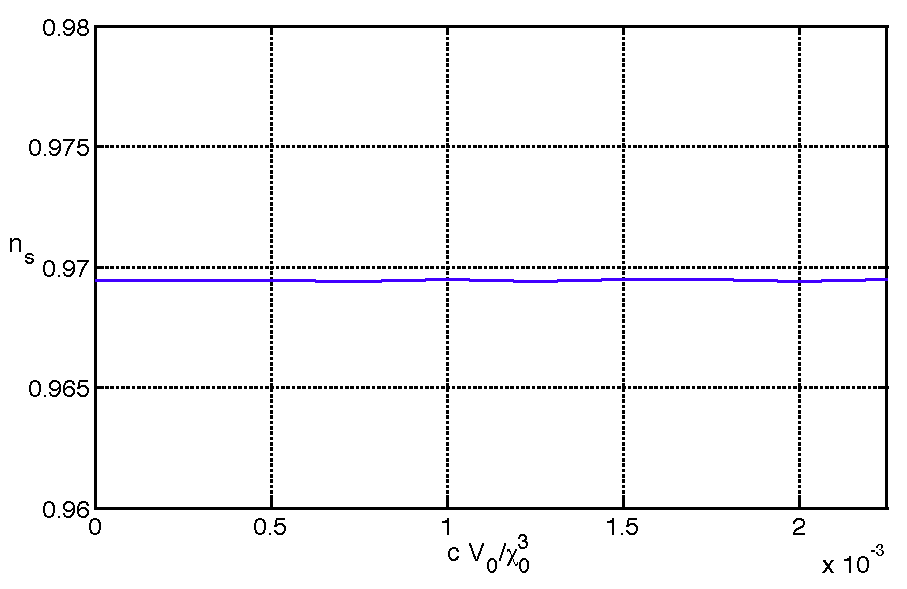}
\includegraphics[width=0.44\textwidth]{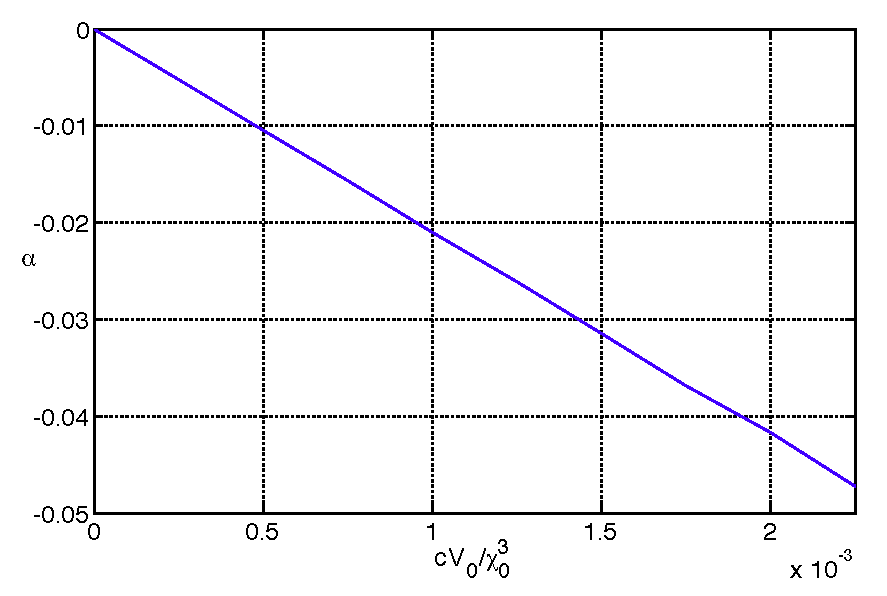} }
\mbox{\includegraphics[width=0.44\textwidth]{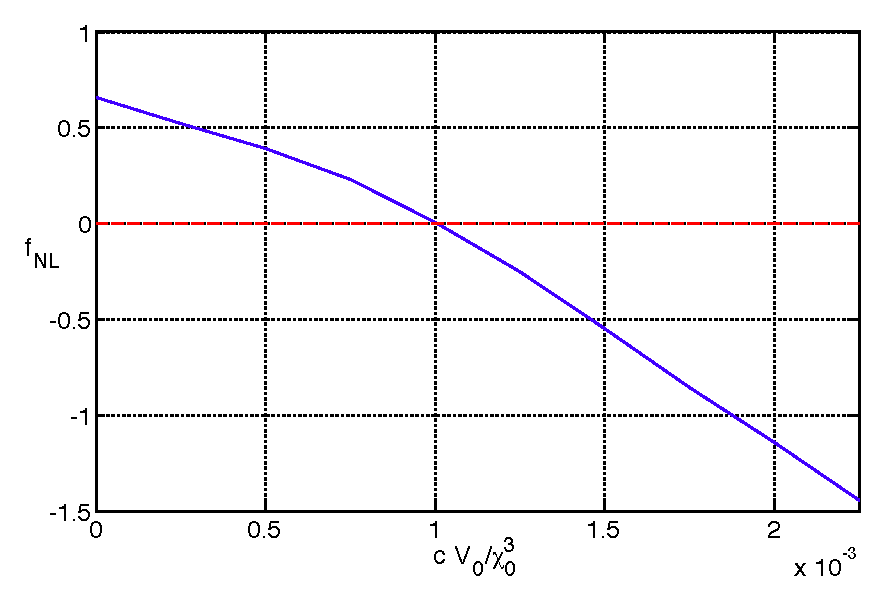}
\includegraphics[width=0.44\textwidth]{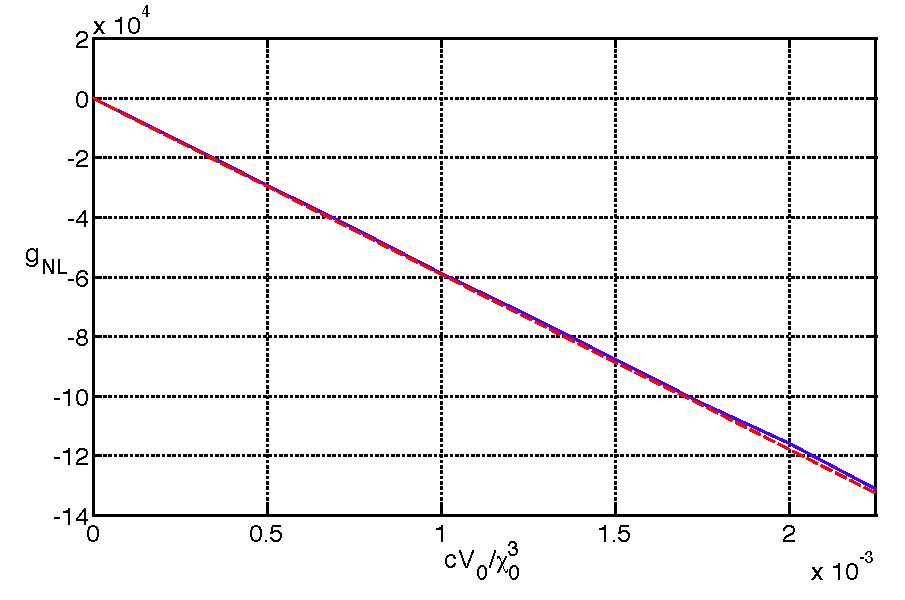} }
\caption{Results for the piecewise isocurvature potential eq.
(\protect\ref{piecewise}). Parametric scan of the pivot scale values
($N=50$) of $n_s$ (top left), $r$ (top right), $\fnl$ (bottom left
solid), the estimate of $\fnl$ (bottom right dashed), $\gnl$ (bottom
right solid) and the estimate of $\gnl$ (bottom right dashed). The
results are shown as a function of the parameter combination $ c V_0
/\chi_0^3$ and measured at the point $N_\phi =N_\chi$.}
\label{fig5}}
\end{figure}

The results for, $n_s$, $\alpha$, $\fnl$ and $\gnl$ at the point the
derivates of $N$ are equal are shown in
Fig.~\ref{fig5} together with the analytical
estimates where appropriate, $r$ always takes  value $r \approx 0.08$.
As above, one can clearly see the correlation between a large
running and the enhanced $\gnl$.

\section{Conclusions}
\label{conclusions}

The unprecedented accuracy of cosmological surveys opens up new
possibilities to probe inflationary physics through correlated
patterns of several observables. After the possible hints of
primordial gravitational waves in BICEP2 data there has been a
significant interest towards running of the spectral index. An
eventual detection of large amplitude primordial gravitational waves would favour a
large negative running to alleviate tension with the accurate Planck
measurements of the spectral index at high multipoles. While the
current BICEP2 data appears compatible with dust \cite{Ade:2014xna} the
possibility for a large tensor to scalar ratio $r$ and large running
$\alpha_s$ is certainly still allowed by observations. The combined
analysis BICEP2 and Planck data and the new data from successors of
BICEP2 is expected to clarify the situation in near future. It is
therefore interesting to investigate the physical ramifications of a
large running in more detail.

In this work we have found new possibilities to discriminate between
inflaton and curvaton fields including through correlated signatures
in scale dependence of the spectral index and non-Gaussian
statistics. In both cases a large running of the spectral index can
be generated by features in the scalar field potential. Using a
simple two parameter description for a local feature either in the
inflaton or curvaton potential we have investigated the
observational signatures allowing for mixed inflaton and curvaton
perturbations. We have shown that a curvaton field generating a
large running of the spectral index also necessarily induces a
specific non-Gaussian signature in the form of an enhanced
trispectrum amplitude $\gnl$ peaked around the feature scale. If the
curvaton induced running is of the same order as the spectral index
$|\alpha_s| = {\cal O}(n_s-1)$ we find the parametric relation
$|\gnl| = {\cal O}(\fnl^2/(n_s-1))$ between the peak values of
$\gnl$ and the bispectrum amplitude $\fnl$. There is no such
amplification if the running is generated by the inflaton(s). Thus
the correlated signatures of running and non-Gaussianities provide a
novel and potentially very useful probe of isocurvature fields
present during inflation. This signature is expected no matter what
the amplitude of gravitational waves.

Another interesting difference concerns the behaviour of the spectrum away
from the feature in cases where a large running is accompanied by
an observable value of the tensor to scalar ratio, $r$.
In the inflaton case a single feature generating a
large running $\alpha_s = -{\cal O}(n_s-1)$ would rapidly lead to
breakdown of slow roll dynamics. A self-consistent setup therefore
requires additional structure which levels out the potential within
$N\sim 10$ e-folds and brings the spectral index closer to unity.
For curvaton induced running due to a single feature, the
inflationary dynamics is not affected and the curvaton can stay in
the vicinity of the feature resulting a decreasing spectral index
over a period up to $N\sim 30$ e-folds. Beyond this regime the
potential either needs to be levelled out by additional structure or
the curvaton will not stay an isocurvature field. If a large running and $r$ are
observed,
therefore, and such a levelling out ruled out within  $N\sim 10$, the origin would have
to be an isocurvature field.

These considerations serve to emphasise the importance of a careful treatment of the second order effects and their correlated behaviour. They could help to uncover potential features and facilitate differentiation between models for the primordial perturbation. As argued here, measuring the running of the spectral index would be very useful in this regard.  Correlated
with the non-Gaussian statistics, the scaling of the spectrum
provides an additional observational tool to discriminate between
curvatons and inflatons. To determine the running more accurately, the widening of the observable window of e-folds would be highly  desirable. Interestingly, such a possibility might be in the offing by future surveys, which will measure 
spectral distortions of the CMB and
probe the spectrum over a range of up to 17 e-folds.

\section*{Acknowledgments}

D.J.M. is supported by a Royal Society University Research Fellowship, and
was supported by the Science and Technology Facilities Council grant ST/J001546/1
during the majority of this work. He thanks Helsinki for hospitality during the
initial stages of this work. SN is supported by the
Academy of Finland grant 257532. KE is supported by the Academy of Finland grants
1263714 and 1218322.

\bibliographystyle{JHEP}
\bibliography{paper}

\end{document}